\documentclass[useAMS,usenatbib]{mn2e}
\usepackage{graphicx}
\usepackage{times}
\begin{document}
\newcommand{\comment}[1]{{\ }}
\newcommand{\degree}{^{\circ}}
\def\Msun{\ifmmode{~{\rm M}_\odot}\else${\rm M}_\odot$~\fi}
\newcommand{\der}[2]{{\frac{d#1}{d#2}}}
\newcommand{\doo}[2]{{\frac{\partial#1}{\partial#2}}}
\newcommand{\fat}[1]{{\bmath{ #1}}}
\newcommand{\FAT}[1]{{\mathbfss{ #1}}} 
\title[Correlation of macroscopic instability and Lyapunov times]{
Correlation of macroscopic instability and Lyapunov times \\ in the general three-body problem
}

\author[S.\ Mikkola and K.\ Tanikawa]
{
Seppo Mikkola$^{1,2}$\thanks{E-mail: seppo.mikkola@utu.fi} and 
Kiyotaka Tanikawa$^2$\thanks{E-mail: tanikawa.ky@nao.ac.jp}
\newauthor
$^1$Tuorla Observatory, University of Turku, 
V\"ais\"al\"antie 20, Piikki\"o, Finland
\newauthor
$^2$National Astronomical Observatory of Japan, Mitaka, Tokyo 181-8588, Japan. 
}

\date{\today}

\pagerange{\pageref{firstpage}--\pageref{lastpage}} \pubyear{}

\maketitle

\label{firstpage}

\begin{abstract}
We conducted extensive numerical experiments of equal mass
three-body systems until they became disrupted. 
The system lifetimes, 
as a bound triple, and the 
Lyapunov times show a correlation similar
to what has been earlier
obtained for small bodies in the Solar System. 
Numerical integrations 
of several sets of differently randomised initial conditions 
produced the same relationship of the instability 
time and Lyapunov time. 
Marginal 
probability densities of
the various times in the three-body experiments are also discussed.
Our high accuracy numerical method for
three-body orbit computations and Lyapunov time determinations
is concisely described.
 \end{abstract}


\section{Introduction}
\cite{LFMlaw} first suggested the existence of a 
relationship
between the local Lyapunov time and a survival time of 
the orbit of an asteroid in the Solar System.
They gave numerical evidence of a powerlaw relation 
\begin{equation}\label{basic}
\frac{t_d}{t_0}=A (\frac{t_e}{t_0})^n,
\end{equation}
where $t_d$ is the survival time, $t_e$ the local Lyapunov time
[which is defined as the inverse of the mean time derivative 
of the logarithm of the variational equation solution during the
interplay period of the system] 
and $t_0$ some normalisation time. 
For the powerlaw index they estimated the value
$n\approx 1.8$.
The relation is by no means an exact one but rather 
the equation fits the centre of a 
scatter diagram of the logarithms of the times.
 The relation have been studied by several authors but
 mainly for asteroid motions and for some artificial mappings.
 Additional discussion  can be found e.g. in 
\cite{Murison} and \cite{LecarARA}.
 \cite{MorbiAndF} discovered two separate regions:
the Nekhoroshev region in which the survival times are
exponentially long, and resonance overlap region in which
an expression like Eq.(\ref{basic}) more accurately apply. However,
they were unable to provide any theoretical explanation.
Later \cite{MurrayH} obtained  evidence against the
generality of the `law' (\ref{basic}).

In this paper we present evidence for a relation of the 
power law type in the general equal mass three-body problem.

\section{Problem and definitions}
The problem studied in this paper is the relationship 
between lifetimes and Lyapunov times of a three-body system.

The lifetime is defined as the duration of the numerical
experiment to the point when the system 
ejects one particle. 
The criterion used was: the single particle is moving away in 
a hyperbolic relative orbit at a distance $>50$ times the semi-major-axis of the 
surviving binary. Then, using the orbital elements,
we compute the (formal) time of pericentre passage 
for this relative hyperbola.
This time, symbol $t_d$, is then considered to be 
the moment of disruption.

In the strict mathematical sense
the Lyapunov time does not exist for unstable triples,
because after the systems disrupt they become
essentially quasi-periodic. We can still use the variational
equations solutions to measure the sensitivity of the system
with respect to variations of initial conditions. 
If the system behaves chaotically, then  $\delta {\fat r(t)}$, 
the solution of the variational equations, has the order of magnitude
 $|\delta {\fat r(t)}|\sim \exp(t/t_e)$. 
Evaluating this at the time $t=t_d$ and solving for $t_e$ gives
(formally) a Lyapunov time as 
\begin{equation}\label{teequation}
t_e=t_d/\ln|\delta {\fat r(t_d)}|.
\end{equation}
Writing the above in another way, one notes that the quantity
\begin{equation}
Z=\ln|\delta {\fat r(t_d)}|=t_d/t_e,
\end{equation}
i.e. the logarithm of the variation at the moment of disruption, 
is actually the lifetime in units
of the Lyapunov time.

\newpage

\section{Numerical method}

\begin{figure} 
\centerline{\includegraphics[angle=0,width=8cm,height=6cm]{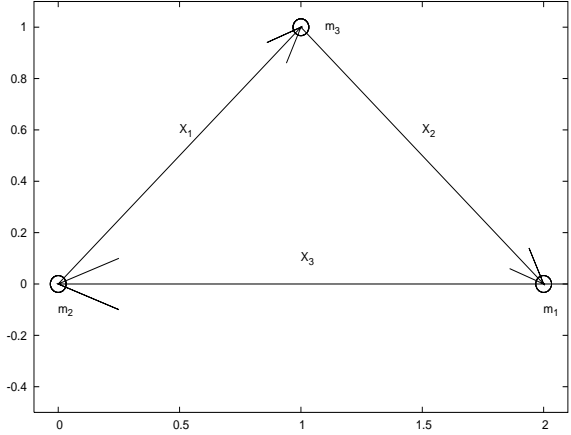}}
\caption {Labelling of vectors in the three-body regularization.} \label{fig-triang} 
\end{figure}

Our numerical method is based on the logarithmic Hamiltonian 
(\cite{logH,Algoreg,PretoTremaine1999,MV60}) which, with the leapfrog
algorithm, gives regular results.
Instead of centre-of-mass coordinates
we use the three
interparticle vectors (see Figure \ref{fig-triang})
\begin{equation}
{\fat X}_1={\fat r}_3-{\fat r}_2;\ \ {\fat X}_2={\fat r}_1-{\fat r}_3;\ \ {\fat X}_3={\fat r}_2-{\fat r}_1.
\end{equation}
as new coordinates.
Let the corresponding velocities be ${\fat V}_k=\dot {\fat X}_k$, then
the centre-of-mass kinetic and potential energies can be written
\begin{equation}
T=\frac{1}{4M}\sum_{i<j}m_im_j {\fat V}_{k_{ij}}^2;\ \ U=\sum_{i<j}\frac{m_im_j}{|X_{k_{ij}}|},
\end{equation}
where $M=\sum_k m_k$ is the total mass and $k_{ij}=6-i-j$. 

The equations of motion are
\begin{equation}
\dot {\fat X}_k={\fat V}_k; \ \ \
\dot {\fat V}_k=-M\frac{{\fat X}_k}{|{\fat X}_k|^3}
+m_k\sum_\nu\frac{{\fat X}_\nu}{|{\fat X}_\nu|^3}
.\end{equation}
and after the application of the logarithmic Hamiltonian modification
they read
\begin{equation}
t'=1/(T+B);\ \ \ \ {\fat X}_k'=\dot {\fat X}_k/(T+B);\ \ \ \ {\fat V}_k'=\dot {\fat V}_k/U.
\end{equation}
where the constant $B=U-T$ is the negative of energy.
The leapfrog algorithm is now possible since 
the right hand sides in 
these equations do not depend on the left hand side variable.

To obtain the solutions to the variational equations 
$\delta {\fat X}, \delta t,\ \delta {\fat V}$ we first
wrote the code to compute the motion and then differentiated that
line by line (i.e. adding the (analytical) code lines for the differentials), 
thus obtaining precise differentials of the numerical
algorithm used. The results are then in the time transformed system
in which time also has a variation. The transformation to physical
system is simple: If we denote the physical system
variations with $\Delta$, we can write for any quantity 
$f$ the equation
\begin{equation}
\Delta f=\delta f-\delta t\: \dot f,
\end{equation}
which gives the desired result.

 Leapfrog is used 
only as a basic method and the results are improved to
high precision by means of the extrapolation method (\cite{BS}). 

The usage of the relative vectors, instead of some inertial
coordinates, is advantageous in attempting to avoid large
roundoff effects. One could also integrate only two of the
triangle sides, obtaining the remaining one from the
conditions
$
\sum_k\fat X_k=\fat 0;\ \ \sum_k \fat V_k=\fat 0.
$
This would hardly reduce the computational effort required
by the method. Instead one may, occasionally, compute the
longest side, and the corresponding velocity, from the above
triangle conditions. Note that the sums of the sides
are not only integrals of the exact solution, but they are also
exactly conserved by the leapfrog mapping. On the other hand,
the extrapolation procedure does not necessarily conserve this 
relation.

The transformation from the variables $\fat X$ to centre-of-mass
coordinates $\fat r$ can be done as
\begin{eqnarray} \nonumber
\fat r_1&=&\frac{(m_3 \fat X_2 -m_2\fat X_3)}{M}\\
\fat r_2&=&\frac{(m_1 \fat X_3 -m_3\fat X_1)}{M} \\
\fat r_3&=&\frac{(m_2 \fat X_1 -m_1\fat X_2)}{M}, \nonumber
\end{eqnarray}
and the velocities obey the same rule.

\section{Initial conditions}
The systems we studied were equal mass systems. We
set all the masses $m_k=1$ (also the gravitational constant
was chosen to be $G=1$).

We attempted to produce random initial conditions although
it is not clear how it should be done since the phase space
is infinite. Initial experiments with various ways showed 
that the results considered in this paper do not depend much 
on how the initial values are randomised. We finally chose
the following algorithm:

First we selected all the 9 coordinates $x_k$ from a uniform distribution
inside a 9 dimensional sphere and similarly the velocity components $v_k$.
Those were reduced to centre-of-mass system and 
scaling factors $c_r$ and  $c_v$ for the coordinates and velocities
respectively were determined from the conditions
\begin{equation}
E=-1,\ \ \rmn{and} \ \  T/U=k,
\end{equation}
where $T$, $U$ and $E=T-U$ are the kinetic energy, potential and total
energy respectively, and $k$, the virial ratio, is a 
selected quantity.
After obtaining the scaling coefficients we simply
replace
\begin{equation}
x_\mu\rightarrow c_r x_\mu,\ \ \ v_\mu\rightarrow c_v v_\mu,\ \ \mu=1,..,9
\end{equation}
 We considered the three values
$k=0$, $k=0.1$ and $k=0.5$. The first one thus gave a system
starting at rest (often called `the free-fall problem') and
thus being a plane problem with zero angular momentum.
The second value $k=0.1$ gives systems with small initial velocities
and the value $k=0.5$ corresponds to systems starting at the
state of virial equilibrium although here it only means 
larger initial velocities and non-negligible angular momentum.

\begin{figure}
\centerline{\includegraphics[angle=0,width=8cm,height=6cm]{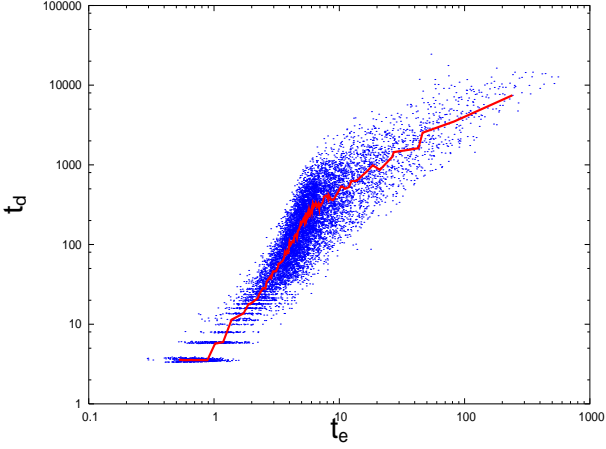}}
\caption{ The scatter diagram of disruption times ($t_d$) and Lyapunov
times ($t_e$) for the free-fall experiments. The line is a median line 
determined such that there
are equally many dots under and above the curve at every value of $t_e$.
Some systematic initial value effects
are visible at small times.
}
\label{scat0figure}
\end{figure}

\section{Results}

\subsection{Experiments}
We conducted 10,000 experiments for each value of the
initial virial coefficient $k=0,\ 0.1,\ 0.5$. Thus altogether
30,000 simulations were run to the point when the system 
disrupted, or the system time exceeded 100,000 time units.
There were only 383 cases in which the computation was halted
because of exceeding the time limit. There was no reason to halt
any of the experiments due to loss of accuracy. Our measure
of accuracy, the ratio of energy error $\delta E$ and the Lagrangian
$L=T+U$ satisfied 
\begin{equation}
|\delta E|/L<1.36\cdot 10^{-11},
\end{equation} 
 in all cases (the number quoted is the maximum), while most values
(in 24350 cases)
were less than $10^{-13}$.

\subsection{Scatter diagrams}

Figure \ref{scat0figure} illustrates the scatter diagram of $(t_e,t_d)$
(times according to Eq.(\ref{teequation}))
in logarithmic scale
for the free-fall experiments. 
The line shown is a median 
produced by sorting the results according to $t_e$, then dividing
the whole range into 50 intervals each containing equally many cases 
for which the median was determined by sorting this group according to
the disruption times $t_d$.

\begin{figure}
\centerline{\includegraphics[angle=0,width=8cm,height=6cm]{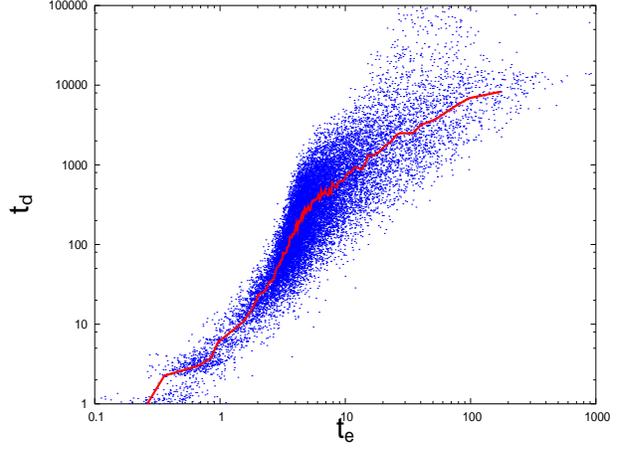}}
\caption{ The scatter diagram of disruption times ($t_d$) and Lyapunov
times ($t_e$) with the median line. 
}
\label{scatfigure}
\end{figure}

Similarly, Figure \ref{scatfigure} illustrates the  $(t_e,t_d)$ scatter diagram
of all the other experiments. The similarity is obvious
although the scatter is somewhat larger.

\begin{figure}
\centerline{\includegraphics[angle=0,width=8cm,height=6cm]{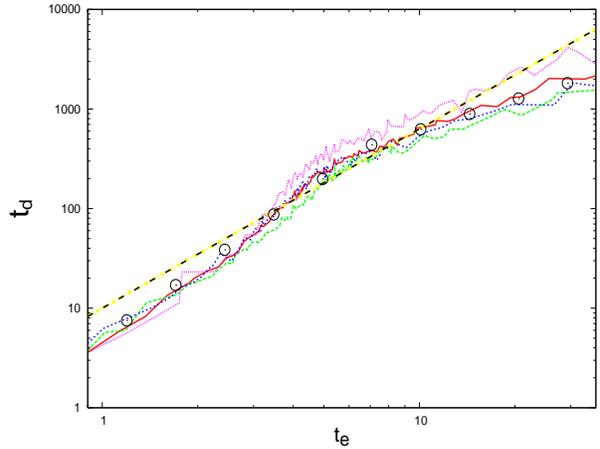}}
\caption{ Median lines for zero, small and large angular momentum systems.
The straight line is for $t_d=10\: t_e^{1.8}$ and the circles for
$t_e=\min(5\:t_e^{2.3},\ 62\:t_e)$. The curve 
above others is for the large angular momentum experiments. The two others,
for zero and small angular momentum, do not differ noticeably.
}
\label{medianfigure}
\end{figure}

A third way of illustrating the correlation of $(t_e,t_d)$ is given in 
Figure \ref{medianfigure} where the medians are plotted in the same figure.
Since in Figures \ref{scat0figure} and \ref{scatfigure} 
most cases are concentrated in a much narrower region than 
the total scale, we restricted this figure into the interval
$.94<t_e<35.2$. These limits were determined such that 
$5\%$ of the cases are in the region $t_e<.94$ and similarly 
$5\%$ in the region $t_e>35.2$. Thus a majority ($90 \%$) of data are included in the
shown interval. Another reason for this restriction is that,
at small times, the systems disrupt almost immediately and
the results are likely to be affected by initial value selection.
The largest times are affected by long ejections without
escape, an effect that may also  partly result from initial
value sampling.

Here three different median curves are produced: 
one for the free-fall experiments,
one for small and medium angular momentum 
and one for large angular momentum.
These were produced by sorting the experiments according to the angular 
momentum value, the smaller half of these were included into the small and medium group and the rest were considered to have large angular momentum.
One can see that the large angular momentum curve is somewhat above the others,
but the difference is minor.
The straight line in the figure represents the relation
$t_d=10 t_e^{1.8},$
which roughly fits the general trend in this region which contains
$90\%$ of the data. On the other hand, the median curves could be
better fitted with two lines: one with $t_d\approx 5 t_e^{2.3}$ for $t_e<6$ and 
with $t_d\approx 62 t_e$ for $t_e>6$, i.e. for large times $t_d\propto t_e $ 
(circles in the figure).

\subsection{Marginal distributions}

\begin{figure}
\centerline{\includegraphics[angle=0,width=8cm,height=6cm]{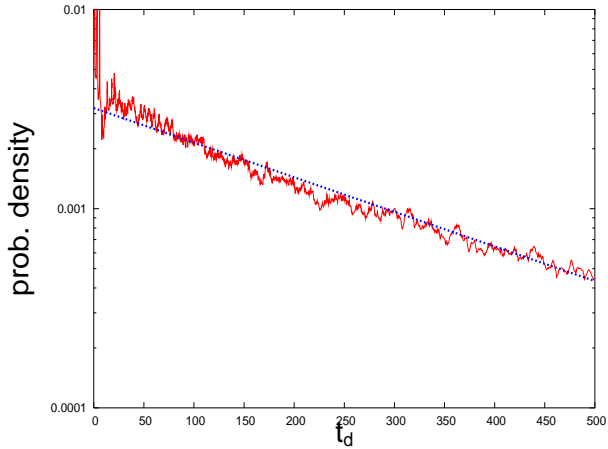}}
\caption{ Probability density of the disruption times. The line 
represents the function $\alpha\exp(-\alpha t_d)$ with $\alpha=1/250$
}
\label{td_psifigure}
\end{figure}

\begin{figure}
\centerline{\includegraphics[angle=0,width=8cm,height=6cm]{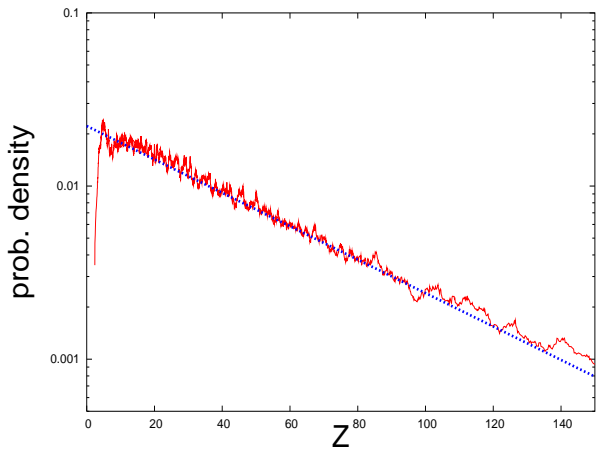}}
\caption{
Probability density of $Z=t_d/t_e$ i.e. the disruption time in units of the
Lyapunov time. The line drawn represents  $\beta\exp(-\beta Z)$ with $\beta=1/45$.
}
\label{X_psifigure}
\end{figure}

\begin{figure}
\centerline{\includegraphics[angle=0,width=8cm,height=6cm]{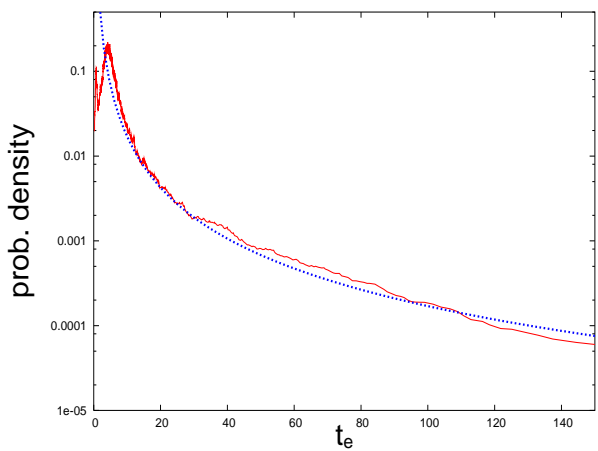}}
\caption{ Probability density of the Lyapunov times $t_e$. 
 This density is clearly not
exponential but, for large $t_e$, is close to the function
$1.8/t_e^2$ drawn in the figure. 
}
\label{te_psifigure}
\end{figure}

It may be of some interest to see what are the marginal distributions
of the various quantities studied here.
We found that, for large values of the times, both $t_d$ 
and $Z=\ln|\delta \fat r|=t_d/t_e$ have exponential probability densities
which are approximately given by
\begin{eqnarray}
\psi(t_d)&\approx& \alpha \exp(-\alpha t_d),\ \alpha=1/250\\
\psi(Z)&\approx&\beta \exp(-\beta Z),\ \beta=1/45. 
\end{eqnarray}
These results are illustrated in the Figures \ref{td_psifigure} and \ref{X_psifigure}.
If one assumes those approximations to be valid for small values also
then it is easy to derive the probability density for $t_e$ as
\begin{eqnarray}\nonumber
\psi(t_e)&=&\int \delta(t_e-t_d/Z)\alpha \exp(-\alpha t_d)\: \beta \exp(-\beta Z)dt_e\: dZ\\&=&\alpha \beta/(\alpha t_e+\beta)^2,
\end{eqnarray} 
which is qualitatively right, as can be seen from Figure \ref{te_psifigure}, where
the probability density of $t_e$ is illustrated. Numerically 
this result is not quite correct, presumably because the
exponential functions for the probability densities of 
$t_d$ and $Z$ are only valid for large values.

\section{Conclusion}
The main result in this investigation is the
obvious existence of a relation of the Lecar-Franklin-Murison
type in the general equal mass three-body problem. 
In the region where most (90\%) of the cases in our diagram
are located the median disruption time can be rather well approximated
by a power law of exponent $n\approx 1.8$ which happens to be
close to what \cite{LFMlaw} obtained in their pioneering paper.
This is somewhat surprising since the systems are totally
different. On the other hand, our results could also be interpreted in the way
that for $t_e<6$ the exponent is $n\approx 2.3 $ and for larger
$t_e$ the exponent is simply $n\approx 1$. If this is the case,
one can conclude that for large $t_e$, i.e. for weakly chaotic
systems, the disruption time is typically proportional
to the Lyapunov time.

\cite{LecarARA} suggested that the relation in Eq.(\ref{basic})
is valid in Solar System in regions where there are overlapping
 mean motion resonances.
However, in the equal mass three-body problem
there are hardly any resonances causing the behaviour, thus the
phenomenon may be more directly related to chaos than to resonances. 
Like \cite{MorbiAndF}, we too do not have any theoretical
explanation for this phenomenon, which seems to be more common
than expected.

In addition to the above, it is interesting to note that (for large times)
the system lifetimes $t_d$ have exponential probability
density both when measured directly in terms of the systems
dynamical time and also when measured
in units of the Lyapunov time  $t_e$. 
A consequence of this is that 
the Lyapunov times have a different probability 
density $\psi(t_e)\propto t_e^{-2}$.
It would be most interesting to find a theoretical 
explanation for these facts, but we are unable to provide one.

\label{lastpage}
\end{document}